\documentclass[
reprint,
 amsmath,amssymb,
 aps,
 prl,
 floatfix,
 superscriptaddress,
 footinbib,
 sort&compress, numbers, merge
]{revtex4-1}

\usepackage{graphicx}
\usepackage{amsmath}
\usepackage{amssymb}
\usepackage{booktabs}
\usepackage{color}
\usepackage[utf8]{inputenc}
\usepackage[tight,sf]{subfigure}	%
\usepackage{natbib}

\newcommand{\be}{\begin{eqnarray}}
\newcommand{\ee}{\end{eqnarray}}

\usepackage[usernames,dvipsnames]{xcolor}

\begin{document}

\title{Curvature-controlled defect localization in crystalline wrinkling patterns}

\author{Francisco L\'opez Jim\'enez}
\thanks{These authors contributed equally to this work.}
\affiliation{Department of Civil \& Environmental Engineering,~Massachusetts~Institute of Technology, 77 Massachusetts Avenue, Cambridge,~MA~02139-4307, USA}

\author{Norbert Stoop}
\thanks{These authors contributed equally to this work.}
\affiliation{Department of Mathematics, Massachusetts Institute of Technology, 77 Massachusetts Avenue, Cambridge, MA 02139-4307, USA}

\author{Romain Lagrange}
\thanks{These authors contributed equally to this work.}
\affiliation{Department of Mathematics, Massachusetts Institute of Technology, 77 Massachusetts Avenue, Cambridge, MA 02139-4307, USA}

\author{J\"orn Dunkel}
\affiliation{Department of Mathematics, Massachusetts Institute of Technology, 77 Massachusetts Avenue, Cambridge, MA 02139-4307, USA}

\author{Pedro M. Reis}
\affiliation{Department of Civil \& Environmental Engineering,~Massachusetts~Institute of Technology, 77 Massachusetts Avenue, Cambridge,~MA~02139-4307, USA}
\affiliation{Department of Mechanical Engineering, Massachusetts Institute of Technology, 77 Massachusetts Avenue, Cambridge, MA 02139-1713, USA}
\date{\today}

\begin{abstract}
We investigate  the influence of curvature and topology on crystalline wrinkling patterns in generic elastic bilayers.  Our numerical analysis predicts that the total number of defects created by adiabatic  compression exhibits universal quadratic scaling for spherical, ellipsoidal and toroidal surfaces over a wide range of system sizes. However, both the localization of individual defects and the orientation of defect chains depend strongly on the local Gaussian curvature and its gradients across a surface. Our results imply that curvature and topology can be utilized to pattern defects in elastic materials, thus promising improved control over hierarchical bending, buckling or folding  processes. Generally, this study suggests that bilayer systems  provide an inexpensive yet valuable experimental test-bed for exploring the effects of geometrically induced forces on assemblies of topological charges.
\end{abstract}

\pacs{46.70.De, 61.72.-y, 46.70.Hg, 89.75.Da}

\maketitle



Topological defects and geometric frustration are intrinsic to two-dimensional (2D) curved crystals~\cite{sadoc2006geometrical}. The minimal number of defects in a periodic polygonal tiling is dictated by Euler's theorem~\cite{Euler}, which relates the surface geometry to the total topological defect charge. The hexagonal soccer-ball tiling is a canonical example, requiring 12  pentagonal defects that are also realized in C$_{60}$ fullerenes~\cite{Kroto:1985aa}. However, 
large 2D crystals often exhibit defect numbers that go substantially beyond the minimal topological requirement~\cite{dodgson1997vortices}. These excess defects  aggregate in molecule-like chains~\cite{irvine2010pleats,bausch2003grain,einert2005grain} that relieve elastic energy costs arising from a mismatch between the crystal symmetry and the curvature of the underlying manifold~\cite{perez1997influence,bowick2000interacting}.   The aggregation of curvature-induced  defects  plays an essential role in various physical processes,  from the classic Thomson problem of distributing discrete electric charges onto a sphere~\cite{thomson1904xxiv} to the assembly of virus capsules~\citep{caspar1962physical,*snijder1993toroviruses} and the fabrication of colloidosomes~\citep{dinsmore2002colloidosomes,*Manoharan2015}, toroidal carbon nanotubes~\cite{liu1997fullerene}, and spherical fullerenes from graphene~\cite{chuvilin2010direct}. Over the past decade, considerable progress has been made in understanding crystal formation in spherical~\citep{bausch2003grain, einert2005grain,wales2006structure,*wales2009defect} and more complex geometries~\citep{giomi2008elastic,giomi2008defective,bowick2004curvature,kusumaatmaja2013defect, bendito2013crystalline, *schmid2014crystalline,vitelli2006crystallography,manoharan2015colloidal}. Yet, empirical tests of theoretical concepts have remained restricted~\cite{bowick2008bubble, irvine2010pleats, irvine2012fractionalization} to paraboloids or mean-curvature surfaces, owing to the lack of  tractable experimental model systems. 

\par
Here, we show through theory and simulations that  curved elastic  bilayer materials offer a promising test-bed for studying defect  crystallography  in arbitrarily shaped 2D geometries.  Building on a recently derived and experimentally validated scalar field  theory~\cite{Stoop15}, we first confirm that thin-film wrinkling reproduces previously established  results for the crystal formation on spherical surfaces~\cite{Brojan14,bausch2003grain}. Subsequently, we demonstrate how curvature and topology determine elastic defect localization on surfaces with non-constant curvature. For typical experimental parameters~\cite{Brojan14}, our analysis reveals the emergence of previously unrecognized robust superstructures, suggesting the usage of topology and geometry to control defect aggregation.

\begin{figure}[b!]
\includegraphics[width=\columnwidth]{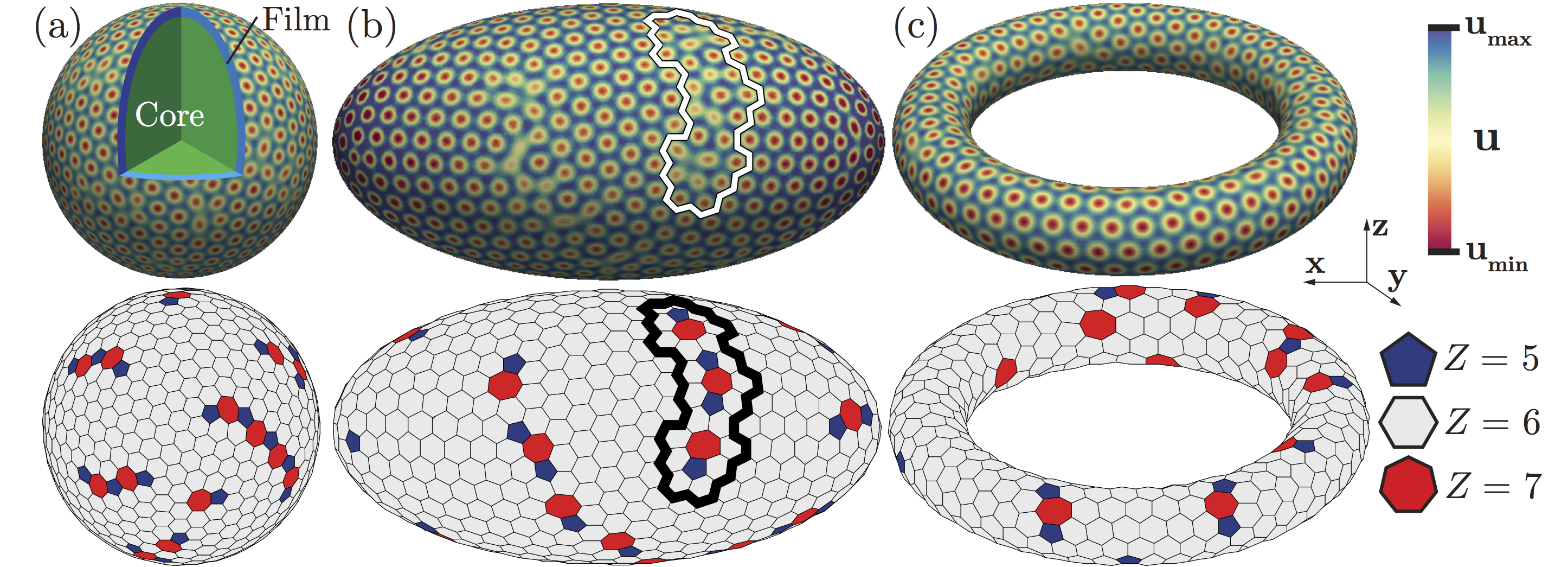}
\caption{(color online) Stress-induced crystalline wrinkling patterns in a thin film of thickness $h$ adhering to a soft core (top), obtained by minimization of Eq.~\eqref{eq:energy}, and their dual hexagonal Voronoi tessellations (bottom) for different surface geometries: (a)~sphere ($R/h = 70$), (b)~ellipsoid ($R_x = 2 R_y = 2 R_z = 110 h$),  (c)~torus ($r/h = 16, R/h = 80$). The color bar represents the surface elevation. The outlined surface domain in~(b) highlights chains of defects. Voronoi cells are color-coded by their coordination number~$Z$.}
\label{fig:Fig1}
\end{figure}

\begin{figure*}[htb!]
\includegraphics[width=\textwidth]{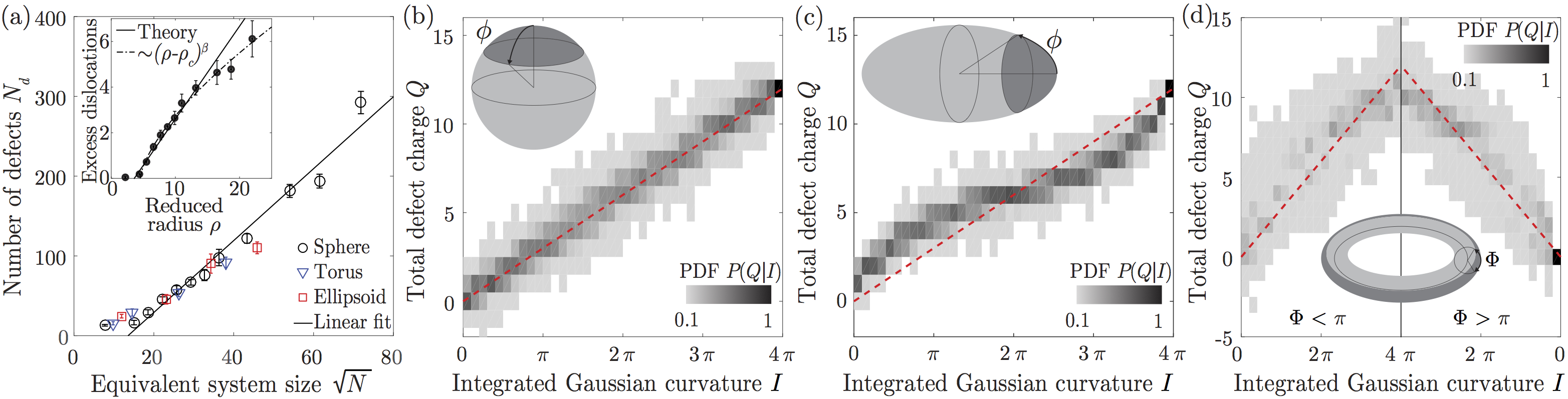}
\caption{(color online) (a)~The total number of defects grows linearly with $\sqrt{N}$, where $N$ is total number of lattice units, exhibiting similar slopes for all geometries (solid line: linear fit for spheres). Inset: Excess dislocations on spheres were predicted~\cite{bausch2003grain} to increase linearly with reduced radius $\rho=R/\lambda$ for colloidal crystals (solid line). For comparison, the best-fit power law $(\rho-\rho_c)^{\beta}$  with $\rho_c=4.5\pm 0.4$ and $\beta=0.67\pm0.08$ is also shown. (b-d)~The total defect charge grows differently with integrated Gaussian curvature $I$ for different geometries. The gray-shading represents the conditional PDFs of the total charge $Q$ for a given value of $I$. The red dashed line corresponds to linear growth $I = (\pi / 3) Q$. Dark regions in the surface sketches  illustrate integration domains.}
\label{fig:Fig2}
\end{figure*}

\par
Our elastic bilayer system comprises a thin stiff film adhered to a soft curved substrate. Recent experiments~\cite{Denis14,Brojan14,Breid13} with spherical substrates showed that, under weak compression, such films can wrinkle into a crystalline dimpled pattern (Fig.~\ref{fig:Fig1}). The experimental patterns are described quantitatively by a generalized Swift-Hohenberg theory~\cite{Stoop15}, which is employed here to obtain predictions for more general geometries. Measuring lengths in units of film thickness~$h$ and focussing on leading-order effects, the theory relates the normal displacement field $u$ of the deformed film to the minimum of the energy functional~\cite{Stoop15}
\begin{align}
\mathcal{E} =& \frac{k}{2} \int_{\omega} d\omega
 \left[ \gamma_0 \left(\nabla u \right)^2
+   \frac{1}{12} \left( \triangle u \right)^2
+ a u^2 + \frac{c}{2} u^4  - \Gamma(u)  \right],
 \label{eq:energy}
\end{align}
where $k=E_f/(1-\nu^2)$, and $E_f$ is the Young's modulus of the film with undeformed surface element $d\omega$. The Poisson's ratio $\nu$ is assumed to be equal for the film and substrate. The nonlinear term $\Gamma(u)= \left[(1-\nu) b^{\alpha\beta}\nabla_{\alpha}u \nabla_{\beta}u + 2 \nu \mathcal{H} (\nabla u)^2 \right] u$ represents stretching forces, with surface gradient $\nabla$ and Laplace-Beltrami operator $\triangle$. The trace $\mathcal{H}$ of the curvature tensor $b^{\alpha\beta}$ defines the mean curvature and $\mathcal{K}$ denotes the Gaussian curvature. Surface wrinkles form when the film stress exceeds the critical buckling stress, corresponding to $a< a_c = 3 \gamma_0^2$ where $\gamma_0$ defines the pattern wavelength $\lambda = 2\pi/\sqrt{6 |\gamma_0|}$, while the amplitude is controlled by~$c$~\cite{Stoop15}.  The curvature-dependent $\Gamma(u)$-term  determines the symmetry of the predicted wrinkling patterns.  In the planar case with $\Gamma(u)=0$, Eq.~\eqref{eq:energy} reduces to the Swift-Hohenberg model 
~\cite{Cross93} and minimization of Eq.~\eqref{eq:energy} produces stripe-patterns. For $\Gamma(u)\ne 0$, the symmetry $u\to -u$ is broken, causing a transition to hexagonal dimple-patterns~\cite{Stoop15}. A systematic derivation~\cite{Stoop15} specified all parameters in Eq.~\eqref{eq:energy} in terms of known material and geometric quantities, so that our predictions can be directly compared to future experiments. Our simulations use the material parameters of Ref.~\cite{Stoop15} for the hexagonal phase with $\lambda\sim 9.1$ throughout.

\par
We analyze Eq.~\eqref{eq:energy} for: spheres of radius $R$, prolate ellipsoids of principal axis $R_x = 2 R_y = 2 R_z$, and tori with aspect ratio $r/R$ between major axis $R$ and minor axis $r$ (Fig.~\ref{fig:Fig1}). For each geometry, we choose $a/a_c \approx 0.98$ and add small random perturbations to the unwrinkled film $u_0\equiv0$. We then minimize Eq.~\eqref{eq:energy} numerically using a custom-made finite element method~\citep{Stoop10,*Vetter13}. From the stationary displacement field,  the dimple centroids are identified as lattice points and the corresponding Voronoi tessellations are constructed (Fig. \ref{fig:Fig1}). We define the topological charge $s = 6-Z$ for each lattice point, where $Z$ is the coordination number. Probability density functions (PDFs) and statistical averages are obtained from simulations with different random initial conditions but otherwise identical parameters.


Since the total number of lattice units $N$ is proportional to the surface area, $\sqrt{N}$ is a geometry-independent measure of the system size. We now characterize the total number of defects $N_d$ and find that it grows linearly with $\sqrt{N}$ for all geometries (Fig.~\ref{fig:Fig2}a), while generally exceeding the Euler bound for the minimal number of defects. For example, we observe exactly $12$ topologically required  defects of charge $s=+1$ (\textit{positive disclinations}) only for small spheres, whereas $N_d$ increasing linearly with slope $m = 4.5 \pm 0.6$ above the critical size $\sqrt{N_c}=14.2 \pm 3.6$ (Fig.~\ref{fig:Fig2}a). In terms of the reduced radius $
\rho=R/\lambda$, $N_c$~corresponds to a critical value $\rho_*  \sim 4.3$, in good agreement with experiments~\cite{Brojan14}.

\par
The increase of $N_d$ with system size can be explained as follows. Each disclination imposes a set change of Gaussian curvature, independently of $\sqrt{N}$. If the mismatch with the substrate's target  curvature becomes too large, additional defects are introduced to screen curvature, thereby lowering the total elastic energy~\cite{bowick2000interacting,travesset2003universality}. To preserve the total charge, 
 excess defects appear as neutral pairs of opposite charge called \textit{dislocations}. For large systems, defects typically form longer chains classified as either neutral \textit{pleats} or charged \textit{scars}~\cite{irvine2010pleats}. For spheres, the number of defects per scar is predicted to grow linearly above $\rho_c$ with slope 
$\approx 0.41 \rho$~\cite{bowick2000interacting,travesset2003universality}. This scaling has been experimentally verified in colloidal crystals~\cite{bausch2003grain}, and agrees with our wrinkling simulations, although a power-law $\sim (\rho-\rho_c)^{\beta}$ with $\rho_c=4.5\pm0.4$ and $\beta=0.67\pm0.08$ also fits our data  well (Fig.~\ref{fig:Fig2}a, inset)~\footnote{We consider a dislocation to be part of a scar when the distance is smaller than or equal to two lattice spacings.}. These findings illustrate that geometry-induced defect formation is insensitive to the details of the lattice interactions~\cite{bowick2002crystalline}, corroborating that  wrinkling patterns provide a viable model to study generic aspects of curved crystals.

\par
The screening effect of dislocations and its dependence on geometry become manifest in the relationship between Gaussian curvature and topological charge~\cite{irvine2010pleats,Brojan14}. The Euler and Gauss-Bonnet theorems connect the sum of topological charges $s_i$ for all elements, $Q = \sum_i s_i$, with the surface integral of the Gaussian curvature, \mbox{$I = \int_A \mathcal{K} dA = \pi / 3 \sum_i s_i = 2 \pi \chi$}, where $\chi$ is the Euler characteristic of the surface. How well this relationship is satisfied over a  \textit{subregion} of the surface provides insight into the geometry dependence of defect localization. For spheres, our results are consistent with Gauss-Bonnet theorem; $Q$ increases linearly with $I$ (Fig.~\ref{fig:Fig2}b). By contrast, for ellipsoidal geometries, $Q$ grows faster than $I$ near the poles ($|\phi| = \pi/2$) and there is an  accumulation of positive charges in high-curvature regions (Fig.~\ref{fig:Fig2}c). Although tori require no topological charge ($\chi = 0$),  our simulations predict the creation of defects that help the dimpled crystal comply with the curved substrate geometry~\cite{giomi2008elastic}. In the outer region of the torus, where Gaussian curvature is positive, we find that $Q$ grows faster than linearly with $I$, which is qualitatively similar to the ellipsoidal case but with larger spread.

\par
\begin{figure}[t!]
\includegraphics[width=\columnwidth]{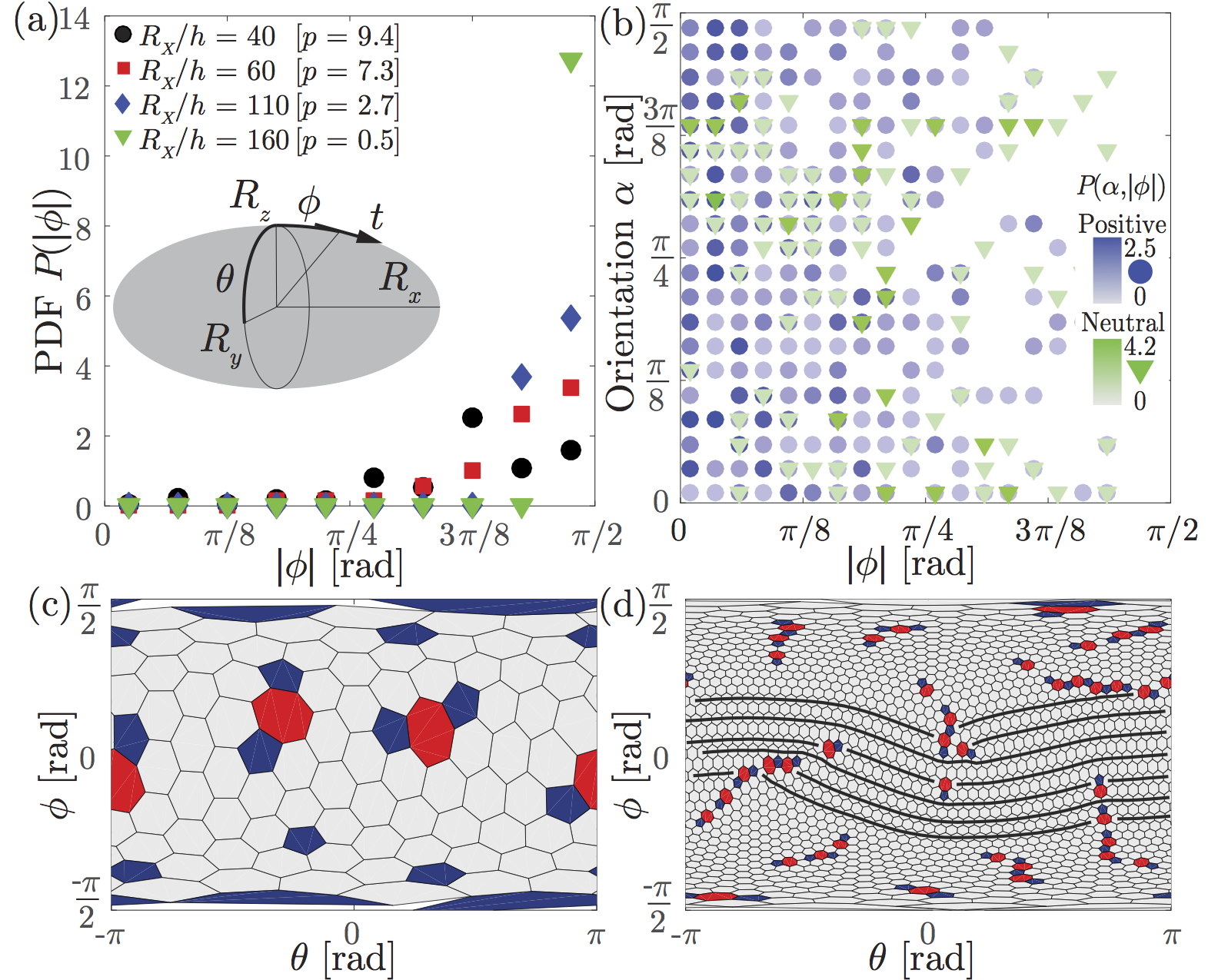}
\caption{(color online) Curvature-induced defect localization on ellipsoids.
(a)~Isolated pentagonal  $+1$-defects accumulate in high curvature regions; $p$  is the average number of single defects per ellipsoid.
(b)~Although defect chains form preferentially near the equator ($|\phi|= 0$) their orientation angles~$\alpha$, measured relative to the tangent vectors $\bf t$, show no significant ordering. (c,d) Voronoi tessellations for $R_x/h=40$ and $R_x/h=160$. Ellipsoids with $R_x/h \ge 110$ show a weak alignment of the lattices along lines $\phi=0$.}.
\label{fig:Fig3}
\end{figure}

Another striking phenomenon is the curvature-induced localization and segregation of oppositely charged defects. For ellipsoids, we find that the PDF for angular position of positively charged disclinations increases strongly towards the poles, where the local Gaussian curvature is large enough to support them  (Fig.~\ref{fig:Fig3}a). With increasing size, additional scars and pleats appear.  Their centroid positions cluster in the equator region around $\phi=0$  (Fig.~\ref{fig:Fig3}b), where the curvature is low and thus cannot support isolated disclinations. To study the orientations of these extended defect \lq molecules\rq, we measure the orientation angle $\alpha$ enclosed by the end-to-end vector $\mathbf{v}$ and the tangent $\mathbf{t}$ along lines with $\theta=const$. We find no significant orientational order for positive or neutral chains~(Fig.~\ref{fig:Fig3}b-d), consistent with earlier simulations based on an inflation packing algorithm~\cite{burke2015role}.

\begin{figure*}[ht!]
\includegraphics[width=\textwidth]{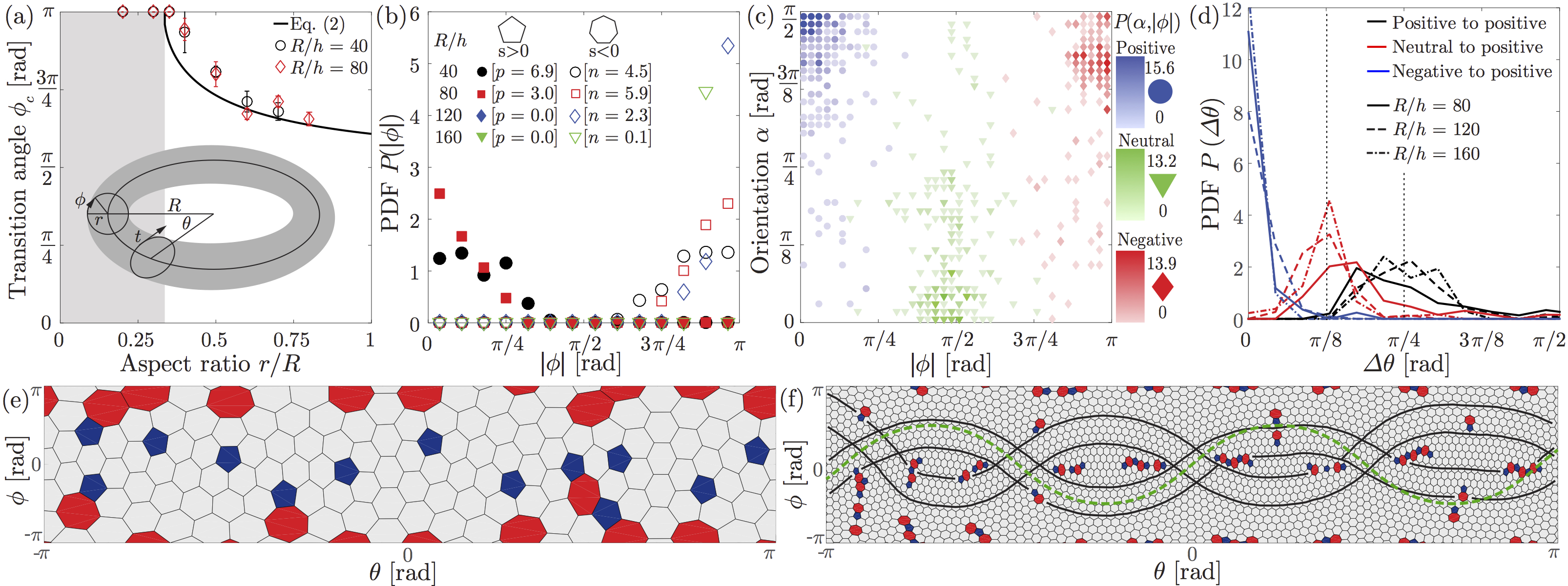}
\caption{(color online) Defect localization and superstructures on tori.
(a)~Pure hexagonal wrinkling phases are stable only for tori with $r/R<1/3$ (shaded). Stripe-like patterns occur at angles $\phi>\phi_c$ for larger aspect ratios. Numerical simulations agree well with the theoretical predictions (Eq. 2, solid line).
(b)~Isolated penta- and heptagonal defects segregate; $p$ and $n$ are the average numbers of positive and negative disclinations per torus.
(c)~Orientations of charged scars and neutral pleats correlate strongly with their transversal centroid position $|\phi|$. (d)~Defect positions are also strongly correlated along $\theta$. (e,f) While this positional ordering is less prominent in small systems [(e), $R/h=40$], it becomes apparent for larger systems [(f) $R/h=160$]. The toroidal superstructure (solid black lines) appears to follow geodesics of minimal integrated Gaussian curvature (dashed green line).}
\label{fig:Fig4}
\end{figure*}

Tori, in contrast to ellipsoids,  contain regions of positive and negative curvature and are more prone to striped wrinkling. To identify the conditions for the pure crystal phase, we recall that hexagonal patterns require $\Gamma(u)\ne 0$ in Eq.~\eqref{eq:energy} whereas local stripe solutions emerge for $\Gamma(u)\to 0$~\cite{Stoop15}. 
The phase boundary can be estimated by first parametrizing the torus using the standard coordinates $(\phi,\theta)$, then assuming a stripe-like wrinkle pattern symmetric along $\theta$ and finally inserting $u(\theta,\phi) \equiv u(\phi)$ into the condition  $\Gamma(u)=0$. Solving for $\phi$, we find the critical transition angle
\begin{align}
\phi_c = \pm \cos^{-1}\biggl[\frac{1-3 \nu }{(5 \nu -1) r/R}\biggr],\label{eq:angle}
	\end{align}
which is independent of the system size and holds as long as $R,r\gg \lambda$. For rubber-like materials ($\nu\approx 0.5$) solutions $\pm\phi_c\in[\frac{\pi}{2},\pi]$ exist for aspect ratios $r/R>1/3$. We thus expect stripe-like wrinkles to dominate near the inner rims of thick tori. For $r/R<1/3$,  $\phi_c$ is imaginary; the symmetry-breaking $\Gamma(u)$-term is globally non-zero, implying a purely hexagonal phase throughout the torus. To verify these predictions, we performed simulations for tori with  $0.2 \leq r/R \leq 0.8$. Defining $\phi_c$ in simulations as the angle beyond which less than half of the wrinkled surface takes the form of hexagonal dimples, we find  good agreement with Eq.~\eqref{eq:angle}, see Fig.~\ref{fig:Fig4}a. The existence of a pure hexagonal phase for $r/R<1/3$ provides a design guideline to study toroidal crystals in future  wrinkling experiments.

\par
Focusing on crystalline patterns on slender tori with $r / R = 0.2$, our simulations show that defect localization is strongly controlled by the interplay of Gaussian curvature and topology. The requirement of a vanishing net charge implies that positive and negative disclinations appear in pairs. Their spatial arrangement can be rationalized with an electrostatic analogy~\cite{bowick2004curvature}, in which defects are interpreted as charged particles and curvature acts as an electric field. In this picture, five-fold disclinations are attracted to regions of positive Gaussian curvature at the outer rim of the torus ($\phi=0$), whereas seven-fold disclinations migrate to the inner region ($|\phi|=\pi$). This geometry-induced separation of charges is directly reflected in the PDF of the individual disclinations (Fig.~\ref{fig:Fig4}b). Analogously to the ellipsoidal case, the total number of isolated disclinations (see  the average numbers $p$ and $n$ for positive and negative charges in Fig.~\ref{fig:Fig4}b) decreases with system size, as defects tend to aggregate in chains (Fig.~\ref{fig:Fig4}e,f). Interestingly, we find that the electrostatic analogy extends to defect chains: positive scars screen Gaussian curvature on the outer rim; negative scars appear in the inner region; and neutral pleats concentrate in the regions of vanishing Gaussian curvature, $|\phi| = \pi/2$ (Fig.~\ref{fig:Fig4}c). Defect chains also become oriented by geometric forces. Measuring the orientation angle $\alpha$ of a chain relative to the tangent vector $\mathbf{t}$ along the $\phi-$direction (Fig.~\ref{fig:Fig4}a), we find that charged scars preferentially align parallel to the equatorial lines such that $\alpha \sim \pi/2$, whereas neutral pleats tend to orient vertically with $\alpha \sim 0$ (Fig.~\ref{fig:Fig4}c). This ordering can be understood qualitatively by considering the end-points of a defect chain. For scars, both end-points have the same charge and therefore migrate to regions of same Gaussian curvature ($\phi \sim 0$ for positively charged endpoints, $\phi \sim \pi$ for negatively charged ones), effectively orienting the scar perpendicular ($\alpha=\pi/2$) to lines $\phi=const$. By contrast, pleats have oppositely charged ends and hence mimic electric dipoles that become oriented by curvature to achieve $\alpha=0$.

\par
Remarkably, our simulations reveal that defects not only orient and segregate in the torus curvature field -- they also break the rotational symmetry of the toroidal crystal in favor of an emergent discrete symmetry. More precisely, by analyzing the lattice structure on large tori, we find that defects arrange along an undulating periodic deformation pattern (highlighted through black lines in Fig.~\ref{fig:Fig4}f). This superstructure still carries a fingerprint of the underlying toroidal geometry: geodesics $g(\phi)=(\phi, \theta(\phi))$ on a torus are solutions of~\cite{oprea2007}
\begin{align}
	\frac{d \theta}{d \phi} = \frac{c\,r}{(R+r \cos\phi) \sqrt{(R + r \cos\phi)^2 - c^2}}.\label{eq:geodesic}
	\end{align}
where $c$ is a constant obeying Claireaut's geodesic relation $c = (R + r \cos\phi) \sin\psi$, with $\psi$ denoting the angle between the tangent $g'$ and $\mathbf{t}$~\cite{oprea2007}. Integrating Eq.~\eqref{eq:geodesic} numerically, we find that the lattice deformation follows a geodesic passing through $\phi=1.68$ at its highest point with $\psi=\pi/2$ (dashed green line in Fig. \ref{fig:Fig4}f). The phase $\theta_0$, found by matching the phase of the geodesic with that of the lattice, varies between samples. The integrated absolute Gaussian curvature along this curve is minimal among all geodesics that do not wind around the $\phi$-coordinate. We therefore hypothesize that alignment along a geodesic of minimal absolute Gaussian curvature yields an energetically favorable crystal structure, for \textit{at least} those lattice parts close to where the geodesic remain nearly straight -- much like wrapping a torus with an inextensible ribbon. If this geometric argument is correct, the superstructure should become independent of the lattice size, as $R/h$ increases. To test this hypothesis, we simulated system sizes $R/h \in\{ 80,120,160\}$, and measured the angular distance $\Delta \theta$ between positive scars near the outer rim.  These simulations indeed reveal a peak in the PDF $P(\Delta \theta)$ near $\pi/4$, which corresponds to one quarter of the geodesic wavelength,  independently of system size (Fig.~\ref{fig:Fig4}d, black curves). Moreover, negative scars on the inside of the torus appear in phase with positive scars ($\Delta \theta=0$), while neutral pleats arrange between scars ($\Delta \theta \sim \pi/8$).
Thus, the geometric lattice superstructure also controls defect-chain localization.

\par
A significant advantage of elastic crystals over fluid-based systems is their fabrication versatility, which enables the exploration of arbitrary geometries and topologies. Our results show that spatially varying curvature can lead to emergent superstructures that determine defect localization. Since defects trigger secondary instabilities~\cite{yong2013prl}, this previously unrecognized phenomenon can be exploited to control hierarchical buckling and folding. More generally, our analysis implies that elastic crystals provide a rich model system for studying the profound interplay between geometric forces and topological charges.

\begin{acknowledgments}
We thank Mark Bathe for sharing computational  facilities. This work was supported by the Swiss National Science Foundation grant No. 148743 (N.S.), by an MIT Solomon Buchsbaum Fund Award (J.D.), by an Alfred P. Sloan Research Fellowship~(J.D.), and by the National Science Foundation, CAREER CMMI-1351449 (P.M.R.).
\end{acknowledgments}

\bibliographystyle{apsrev4-1}
\bibliography{references}

\end{document}